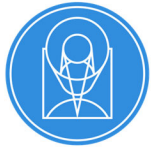

# JWST TECHNICAL REPORT

| Title: Empirical Modeling of Zodiacal Backgrounds to Improve JWST NIRISS/SOSS Data Reduction | Doc #: JWST-STScI-009046, SM-12<br>Date: 10 June 2025<br>Rev: A |
|---|---|
| Authors: Tyler Baines, Phone:<br>Aarynn Carter, Néstor Espinoza, Kevin Volk, Joseph Filippazzo, Loic Albert | Release Date: 4 September 2025 |

## 1 Abstract

Zodiacal light—arising from both the thermal emission and scattered sunlight by interplanetary dust—is the dominant component of the sky background in NIRISS Single Object Slitless Spectroscopy (SOSS) observations. The GR700XD grism disperses the zodiacal background across multiple diffraction orders, producing a characteristic two-dimensional, spectral order-dependent background pattern on the detector that reflects the combined contribution of overlapping orders. Observations reveal significant variability in background intensity driven by JWST's sky pointing and seasonal changes, underscoring the need for precise background subtraction during SOSS data reduction. Current methods rely on a generic background template derived during commissioning, scaled to match individual exposures. However, mis-scaled templates can leave behind structured residuals that may compromise the precision of exoplanet transit depth measurements. To improve background modeling, we conducted two calibration programs (PID 4479 and 6658) using the FULL frame readout mode and a 5-row by 2-column mosaic pattern to sample a range of sky positions. These observations enable empirical reconstruction of the sky background and provide detailed insights into its spatial and spectral characteristics. We present a library of empirically derived background templates and evaluate their performance, alongside the current template, by benchmarking against the PID 2113 dataset, which includes contemporaneous background exposures. This work supports aims to enhance background subtraction for SOSS time-series to achieve higher-precision exoplanet spectroscopy with JWST.

## 2 Introduction

The James Webb Space Telescope (JWST) is designed to enable transformative science, including the atmospheric characterization of terrestrial exoplanets. One of its key observing modes, the Single Object Slitless Spectrograph (SOSS) mode of the Near Infrared Imager and Slitless Spectrograph, provides high-precision spectrophotometric time-series observations of transiting exoplanets. Accurate calibration of NIRISS/SOSS data is essential for reliably extracting the planet's signal and maximizing scientific return. This calibration must address various instrumental and observation effects that impact data quality, including accurate subtraction of the sky background.

Background subtraction poses particular challenges for slitless spectroscopy. Unlike slit spectroscopy or imaging, all sources in the field—including the sky background—are dispersed by the grism, leading to signal overlapping and blending, and a more complex two-dimensional background structure. The sky background in JWST data arises from both observational and instrumental sources. including, zodiacal light, interstellar medium emission, stray light within the observatory, and thermal emission from JWST's own mirrors (see Rigby et al. 2023 for detail breakdown).

Zodiacal light, originating from interplanetary dust that scatters sunlight and emits thermally, is the dominant contributor to the sky background in NIRISS/SOSS observations. The GR700XD grism disperses this background across multiple diffraction orders, producing a spectral order-dependent background pattern on the detector with a prominent discontinuity where the zeroth-order projection ends. This shape is understood to arise primarily from diffraction of order 1 and 2 on the left side of the detector, and from orders 0, 1, and 2 on the right, with minor contributions from order –1 and 3.

Before launch, the background structure was assumed to be stable and scalable in intensity. However, early reductions of science data revealed this assumption to be incorrect. The current background model—developed during commissioning—often leaves significant residual structure particularly discernible near the discontinuity (see [NIRISS SOSS Recommended Strategies](#), 2024). These issues are exacerbated when contaminating traces from other sources prevent the use of clean regions for template scaling, highlighting the need for a more flexible, empirical approach to background modeling.

This report characterizes the zodiacal light background in NIRISS/SOSS observations to improve background subtraction during data reduction. We begin by investigating the physical origin of the background using simulated data, revisiting pre-launch assumptions and identifying sources of variability (Section 3). We then construct empirical background templates using data from dedicated calibration programs and supplementary datasets, outlining the processing methodology (Section 4). These empirical backgrounds are analyzed across multiple sky positions to examine spatial and spectral variability (Section 5), followed by a study of seasonal trends using repeated

Check with the JWST SOCCER Database at: https://soccer.stsci.edu
To verify that this is the current version.
2

observations of BD+60 1753 (Section 6). We assess the performance of these templates through background subtraction tests and quantify improvements in data quality (Sections 7–8). We conclude by summarizing our findings and offering practical recommendations for users (Section 9).

## 3 Investigating the Sky Background in SOSS Observations

The unique shape of the SOSS sky background exhibits several key characteristics resulting from the grism dispersing an otherwise uniform illumination across multiple diffraction orders, as illustrated in Figure 1. The background intensity gradually increases along the dispersion axis (left to right on the x-axis) until around pixel column 700, where a sharp discontinuity is observed. This transition corresponds to the region where the zeroth-order contribution diminishes, and only higher diffraction orders remain. (For reference, this pixel location aligns with a wavelength of roughly 2.1 $\mu$m in the first-order spectrum, were it present.) The discontinuity arises from the bright, compact nature of the order 0 core, in contrast to the broader, more diffuse high-order traces. Its location is determined by the size of the pick-off mirror (POM) and the offset between the direct image position and the projected zeroth-order spectrum. While the presence of this feature was understood early on in JWST operations, its precise magnitude and spatial structure remain complex, influenced by factors such as the grism-induced displacement of the stellar image on the detector.

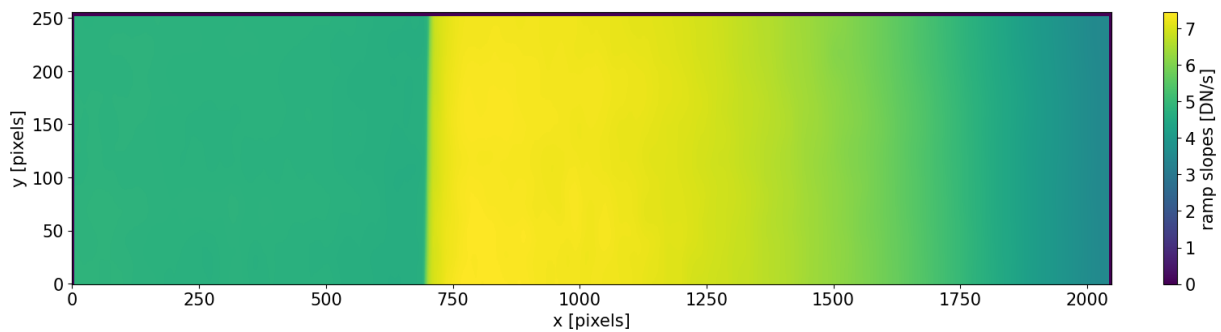

*Figure 1. shows the current background template derived during commissioning for the SUBSTRIP256 to illustrate the SOSS background shape dominated by zodiacal light. The discontinuity is observed around pixel column 700.*

We began with an existing Interactive Data Language (IDL) simulation script that generates a general approximation of the zodiacal background (L. Albert). As a seed, we also made use of a full 2D image simulation of the SOSS curved, defocused traces produced for diffraction orders –1 to 3, as generated by the IDTSOSS simulator (see Section 9 of Albert et al. 2023) using the expected zodiacal light spectrum as the input. To produce the background model, the IDL script shifts this seed 2D image and adds up the flux for all permitted source positions in the field of view, essentially limited by the size of the POM—roughly a square centered around the detector, oversized by about 140 pixels all around (see Figure 2 of Albert et al. 2023). Figure 2 illustrates conceptually how the IDL script works.

The IDL code was converted to Python and we recreated the zodiacal background model with good agreement. Building upon this foundation, we modified the code to produce background images of the individual diffraction orders -1 to 3. The simulated background for each diffraction order and their combined sum produces a result that qualitatively reproduces the general background structure, replicating the sharp discontinuity pixel around column 700 as illustrated in Figure 3. Note that this simulation was performed using the FULL frame detector field of view, whereas Figure 1 shows the SUBSTRIP256 subarray. The agreement is qualitative but provides insight into the origin of the structure seen in the subarray data.

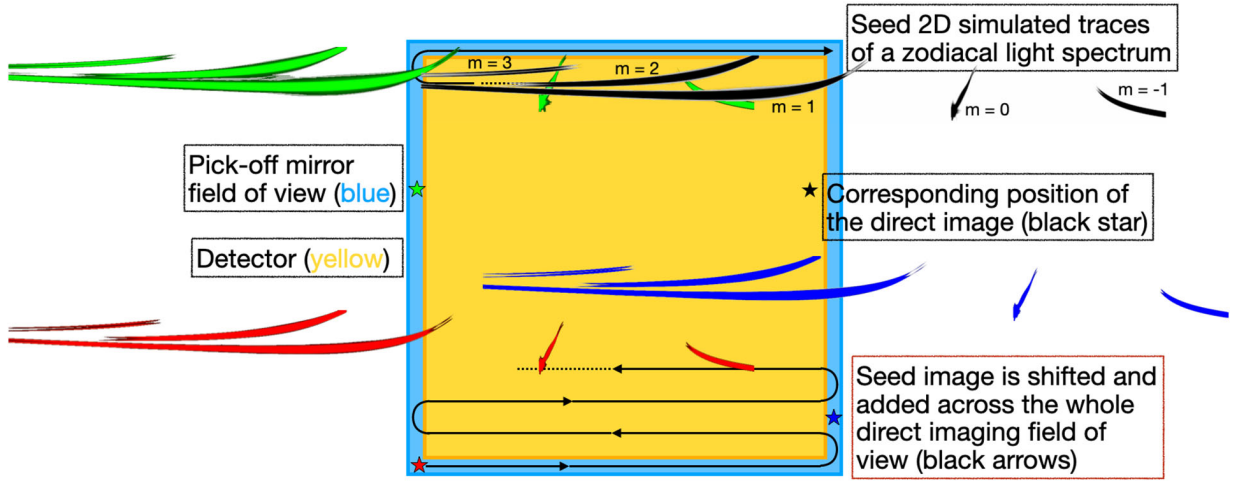

*Figure 2. Zodiacal background simulation using a seed 2D spectral trace (orders −1 to 3) shifted across the pick-off mirror's field of view. Due to vertical cross-dispersion, sources below the red star do not contribute, producing the dark lower region. Sources left of the red and green stars fall outside the field of view, placing order 0 near x ≈ 700 and creating the vertical discontinuity. Orders 1 and 2 remain across all x-positions, contributing to the relatively constant background.*

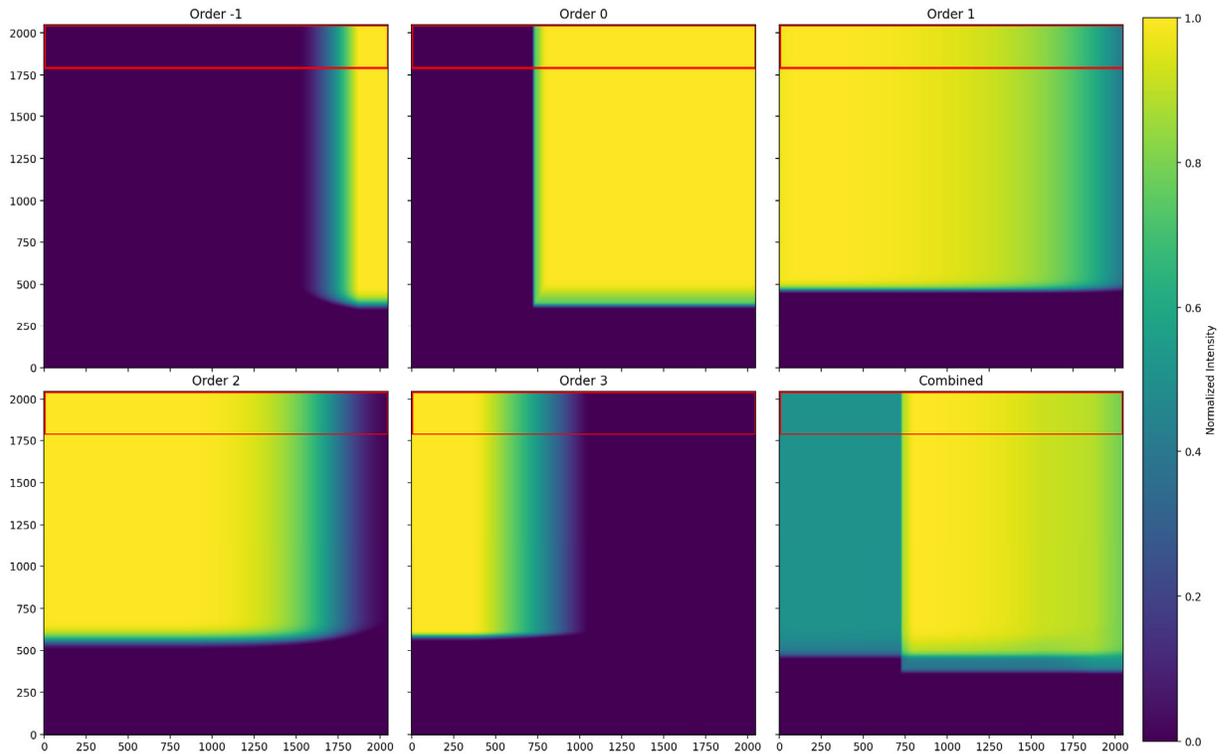

*Figure 3. Simulated FULL-frame background contributions from five diffraction orders using our toy model, designed to replicate the spatial structure of the SOSS background. Each panel shows the normalized intensity map of the background signal associated with a specific spectral order. The final panel presents the combined background obtained by summing the individual contributions. The red box outlines the SUBSTRIP256 subarray. This simulation demonstrates that overlapping contributions from higher and lower orders significantly shape the overall background structure.*

## 4 Empirical Derivation of SOSS Background Templates

Accurate modeling of the sky background is vital for achieving high-precision spectrophotometric time-series measurements with NIRISS/SOSS. To support this, we adopt an empirical strategy that involves collecting dedicated background exposures and using small telescope dithers to shift astrophysical sources across the detector. Because the zodiacal background remains fixed in detector coordinates while the sources move, this technique allows us to disentangle the diffuse background signal from localized source contamination. The resulting clean background data form the basis for constructing high-fidelity templates that account for spatial and seasonal variations, enabling more effective background subtraction during data reduction.

4.1 Data Collection and Filtering

To construct robust empirical background templates, we conducted two dedication calibration programs (PID 4479 and 6658) using the FULL-frame readout mode with the CLEAR/GR700XD configuration. These programs targeted multiple sky pointings near the ecliptic plane and at ±30° in ecliptic latitude to sample a broad range of zodiacal light conditions. Each pointing was observed using a 5-row by 2-column mosaic pattern with 75% overlap in both x and y directions, producing 10 exposures per field. By systematically sampling different regions of the detector, this strategy improves spatial coverage and enables a cleaner, more detailed reconstruction of the underlying sky background. A few exposures from PID 6658 (e.g., observations 004, 007, 008, and 009) failed resulting in 8–9 usable exposures, though this did not impact our ability to extract reliable backgrounds.

We retrieved the Level 1 rate products—detector-corrected images calibrated/reduced with the JWST pipeline—from the Mikulski Archive for Space Telescopes database for all exposure associated with these programs. Additionally, we incorporated any other available exposures from unrelated SOSS programs also read out in the full frame read-out mode that may be suitable. Figure 4 illustrates the sky distribution of these calibration programs in ecliptic coordinates, providing context for the sky coverage and relevance of the background dataset. For reference, we also show the pointings for planned Cycle 4 background observations (PID 9284).

To prepare the background dataset for processing, we first grouped exposures based on the programs, sky position, and observation identifier. In some cases, exposures within a single observation were acquired several days apart (e.g., observation 007 from program 4479 and observation 005 from program 6658). To accommodate these time-separated sequences, we assigned a numerical tag to each subset of exposures acquired within a one-day interval and treated them as an independent background set.

To ensure adequate sampling for empirical reconstruction, we required a minimum of four exposures per set—consistent with the intended mosaic pattern strategy. After filtering, we retained 17 discrete background sets for subsequent processing. Table A-1 in the appendix provides a listing of these grouped background datasets use for empirically reconstructing the background.

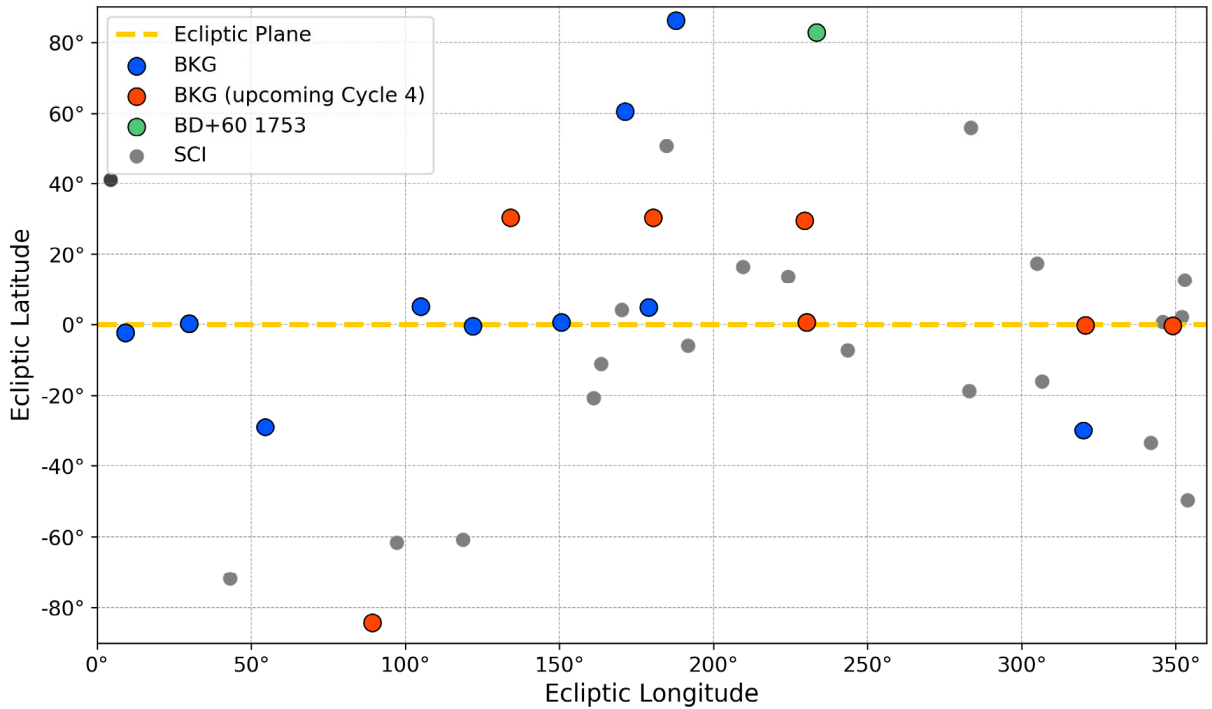

*Figure 4. Spatial coverage of sky pointings for dedicated background observations. Also shown are the pointings of relevant science observations and upcoming Cycle 4 background programs (PID 9284). The ecliptic coordinates of BD+60 1753 are indicated (green), as this target's repeated observations make it well-suited for studying seasonal background variations.*

4.2 Data Processing: Empirical Background Extraction

Prior to extracting an empirical background, we apply the flat-field correction (jwst_niriss_flat_0275.fits) to each image in a background set to mitigate pixel-to-pixel sensitivity variations that may introduce artifacts in the reconstruction of the background. This initial flat-fielding step is intended solely to improve the quality of the background model and is reversed following the background extraction step. To extract the background from each background set, we begin by masking astrophysical sources—primarily dispersed spectral orders 0, 1, and 2—to prevent contamination of the background estimate. As a first approximation, we compute the median across the image stack to generate a coarse background model, which serves as input for source detection using Photutils (`detect_threshold` and `detect_sources`). For source detection, we identify pixels exceeding a $2\sigma$ threshold above the background above the initial background. To improve the fidelity of the source mask, we apply a binary dilation (`binary_dilation` from `scipy.ndimage`) to expand the masked regions and capture adjacent pixels associated with source flux. We also exclude pixels flagged with non-zero data quality (DQ>0) values, combining them with the source mask.

With the source and flagged pixels masked, we recompute the background using a sigma-clipped median (`sigma_clipped_stats` from `astropy.stats`: Astropy Collaboration. 2025), producing a cleaner and more robust background image. Any remaining NaN pixels in the resultant background image are filled using `maskfill` (van Dokkum & Pasha 2024), which infers missing values from surrounding pixels where edge pixels are used to extrapolate inwards to fill masked regions. As a final refinement, we apply a non-local means denoising algorithm implemented in the scikit-image library (Darbon et al. 2008) to smooth residual noise while preserving fine-scale structure and the discontinuity. At this stage the flat-fielding is reversed and added back to the resulting empirical background. This procedure is applied independently to each of the 17 background sets, yield 17 high-fidelity empirical background models.

## 5 Variability in Empirical Backgrounds Across Sky Pointings

With the empirical backgrounds extracted, we now assess how the background levels and spatial structures vary across sky pointings. This variability is crucial to understanding the robustness and limitations of background subtraction strategies. Figure 5 displays the SUBSTRIP256 subarray cutouts for all 17 empirical backgrounds, revealing differences in absolute levels between pointings while the spatial structure remains consistent. To quantify this difference, Figure 6 plots the median background level of each empirical template as a function of absolute ecliptic latitude.

To explore the spatial structure in more detail, Figure 6 presents side-by-side comparisons of empirical backgrounds and simulated zodiacal light spectra for each sky pointings. The **left column** of Figure 7 shows cross-section profiles taken from row 225 in the SUBSTRIP256 array (equivalent to rows 1791-2047 in the FULL array, Python zero-based indexing) of each empirical background. We include both linear-scale (top-left) and normalized (bottom-left) versions of the profiles. Each profile is normalized by its median value left of the discontinuity. When normalized this way, cross-sections exhibit minimal variation left of the discontinuity, but clear modulation is observed on the right side, which is strongly influenced by the dispersed zeroth-order flux. This indicates that the order 0 does not scale uniformly with orders 1 and 2 across different pointings—while the normalized profiles align well where higher orders dominate, the divergence on the right side reflects variations in the relative zeroth-order contribution.

To investigate the origin of the modulation, we generated model zodiacal light spectra using the JWST Background Tool for each sky position in the empirical backgrounds. These are shown in the right column of Figure 7. The top-right panel shows surface brightness versus wavelength; the bottom-right panel shows the same spectra normalized at 1.3 µm—the peak response of Order 1, which dominates the region left of the discontinuity. The models diverge beyond ~3 µm, where thermal emission from interplanetary dust becomes significant. This variation plausibly explains the modulation seen in the zeroth-order region. Order 0 spans 0.6–5.0 µm and is more sensitive to long-wavelength changes, while Orders 1 and 2 primarily cover 0.6–2.8 µm.

While this interpretation is qualitatively consistent with observed trends, several caveats remain. While this interpretation is qualitatively consistent with the observed trends, several caveats remain. Firstly, Order 0 is not the sole contributor to the right-hand side of the detector; higher diffraction orders also influence the total background signal and its spectral shape. This is supported by throughput measurements for spectral Orders 0 through 3 (see Figure A-1 in the Appendix). Secondly, there are known mismatches between zodiacal light continuum models and imaging observations from NIRISS and NIRCam at shorter wavelengths (internal investigation ongoing). A full investigation of the physical mechanisms responsible for this spectral diversity—such as dust composition, distribution, or viewing geometry—is beyond the scope of this study. However, the current zodiacal light model still provides a useful proxy for interpreting SOSS background behavior and its sensitivity to sky position.

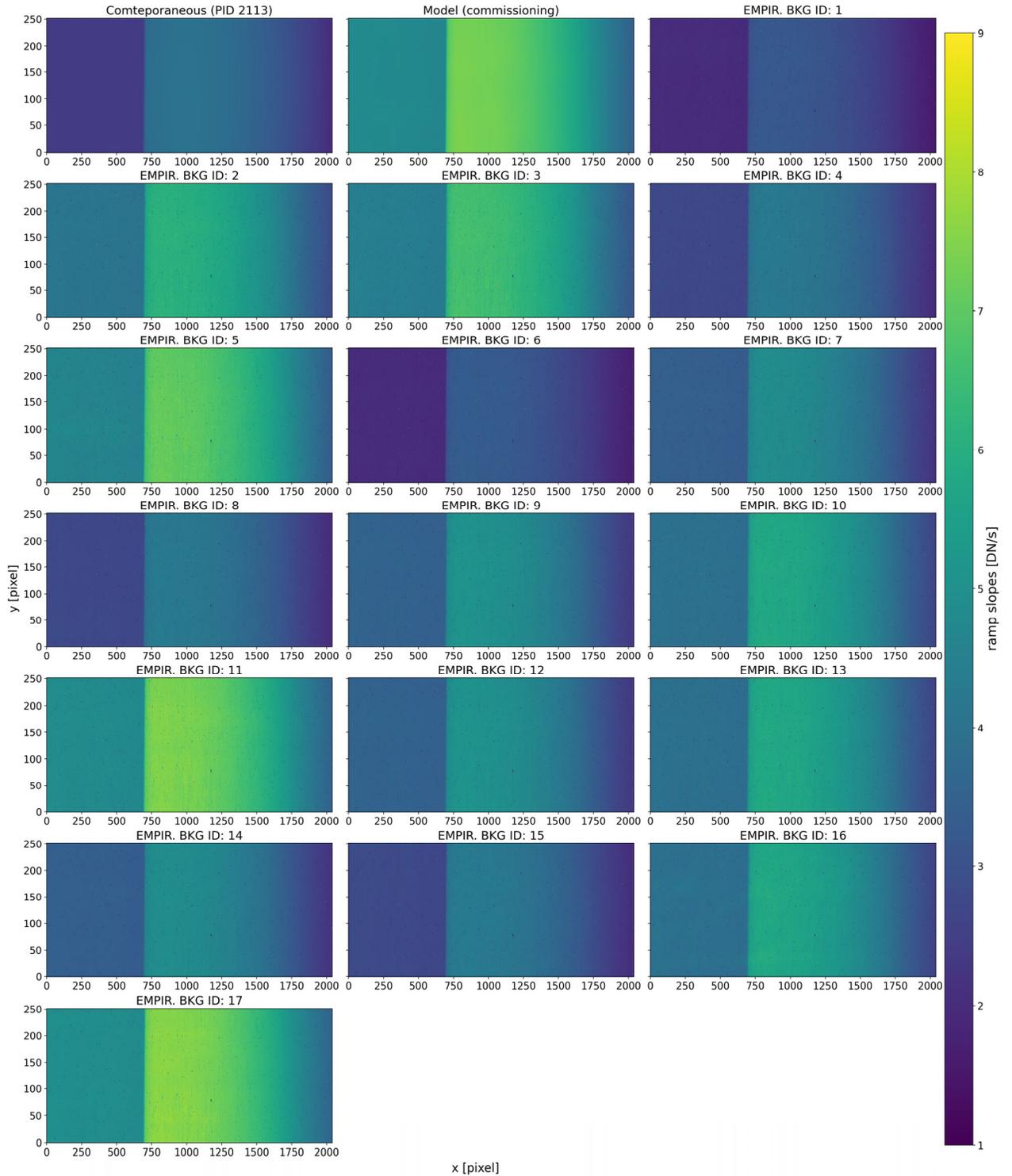

*Figure 5. Empirical background templates derived from stacks of background mosaic exposures taken at different sky positions, illustrating spatial structure and variation in background levels across the SUBSTRIP256 subarray. All images are displayed with the same color scale (units of DN/s). For comparison, the contemporaneous background from the science observation (PID 2113) and the current reference model from commissioning are also shown.*

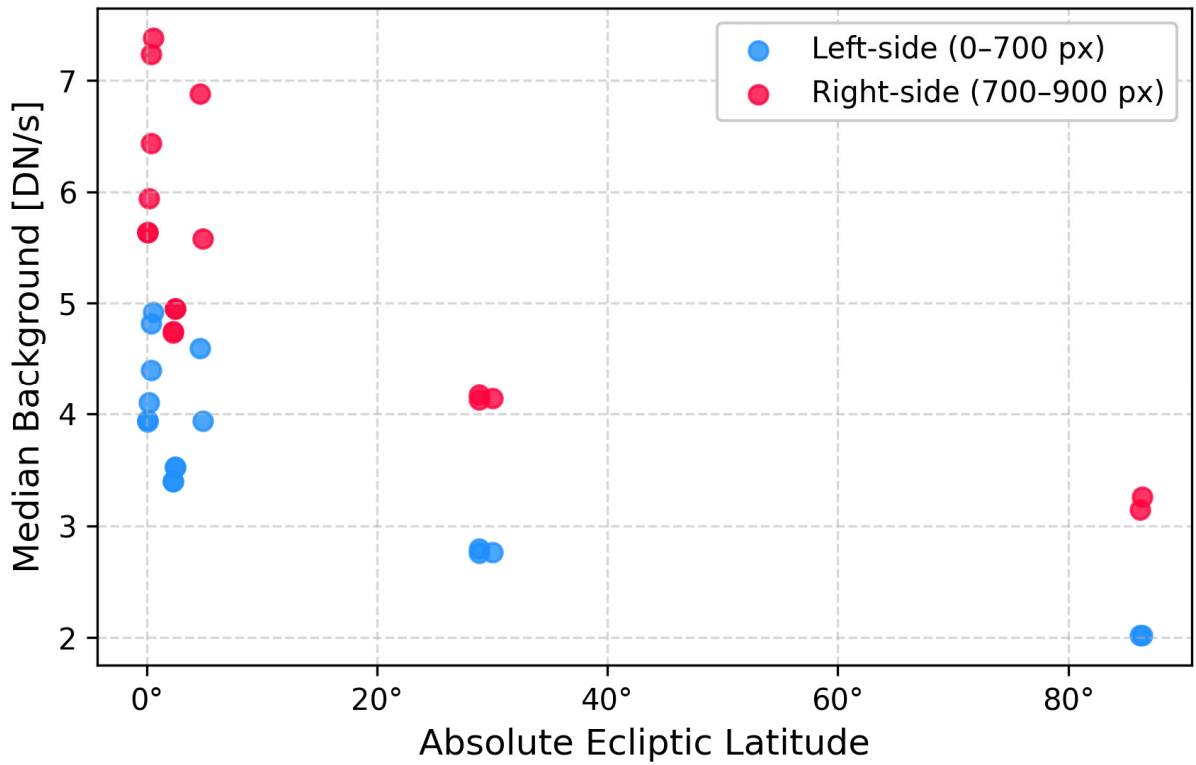

*Figure 6. Measured median background level, left and right of the discontinuity, as a function of absolute ecliptic latitude for the 17 empirical backgrounds images.*

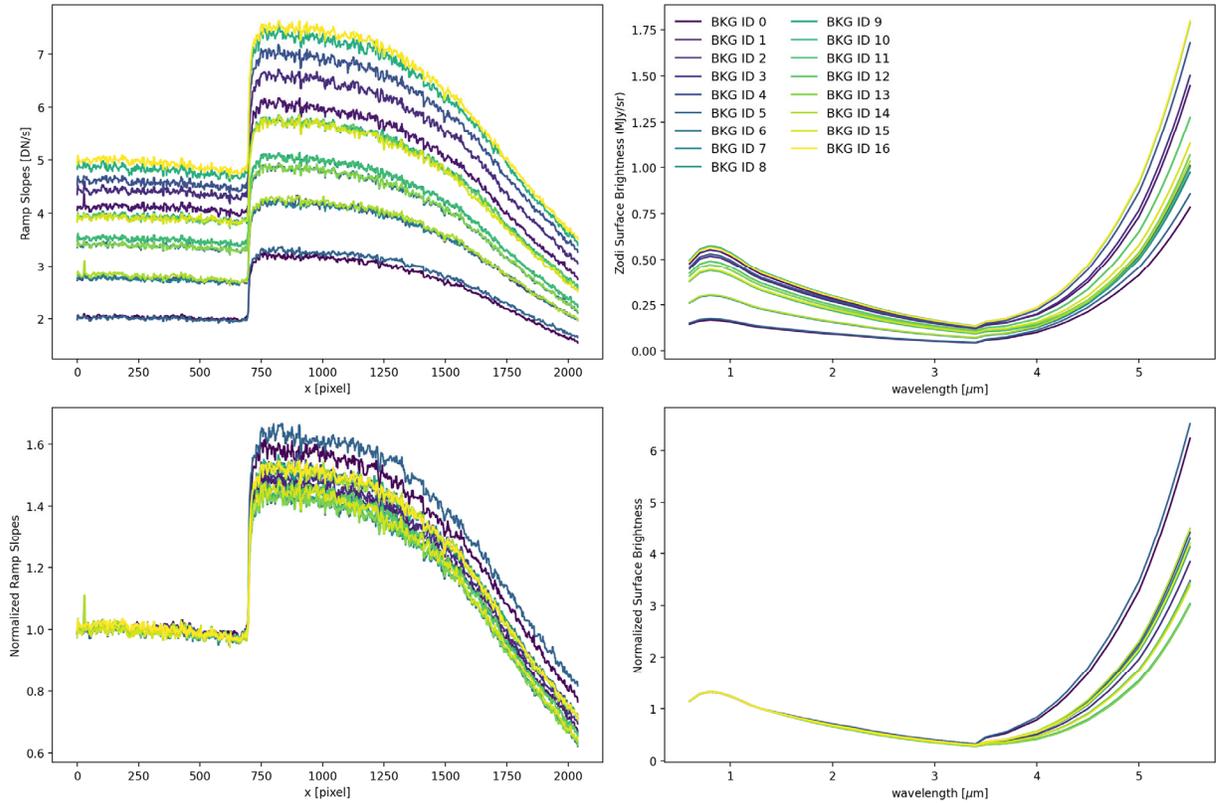

*Figure 7. Spatial and spectral characterization of the empirical background library. Top-left: Row 225 cross-section profiles from all 17 backgrounds, median-filtered to suppress noise, showing variation across SUBSTRIP256. Top-right: Simulated zodiacal spectra from the JWST Background Tool at matching sky positions. Bottom-left: Same profiles as top-left, normalized by the median left of the ~750-pixel discontinuity to highlight variability in the 0th-order region. Bottom-right: Same spectra as top-right, normalized at 1.3 µm. Increased variation beyond 3 µm may explain modulation in the 0th-order region due to changing zodiacal emission.*

## 6 Seasonal Variations of the Sky Background

While dedicated calibration programs have successfully mapped spatial variations in the SOSS sky background—sampling regions above, below, and near the ecliptic—probing seasonal variability poses greater challenges. Repeated background observations at fixed sky locations require significant telescope time and yield limited direct science return, making such programs difficult to justify.

Fortunately, archival datasets offer an alternative. In particular, repeated calibration observations of the A-type standard star BD+60 1753—originally intended for flux monitoring, wavelength calibration, and sensitivity tracking—provide a valuable opportunity to investigate seasonal background trends. These observations, conducted over nearly three years using the SUBSTRIP256 subarray, sample a consistent sky location at different times of year.

To analyze seasonal behavior, we retrieved Level 1 rate products for all BD+60 1753 observations. Source flux was masked using a column-wise median thresholding algorithm, and the background was estimated by computing the sigma-clipped median from pixels to the left of the zodiacal discontinuity ($x < 700$). Figure 8 shows the measured background levels as a function of time, folded into a single year to reveal seasonal patterns. Despite BD+60 1753's high ecliptic latitude—where background levels are typically low—we observe up to ~42% variation (~1.0 DN/s) across the year, suggesting that even high-latitude targets can be affected by seasonal changes in the thermal sky background. The gap between days ~280–365 reflects the target's annual solar avoidance period, during which it is unobservable by JWST.

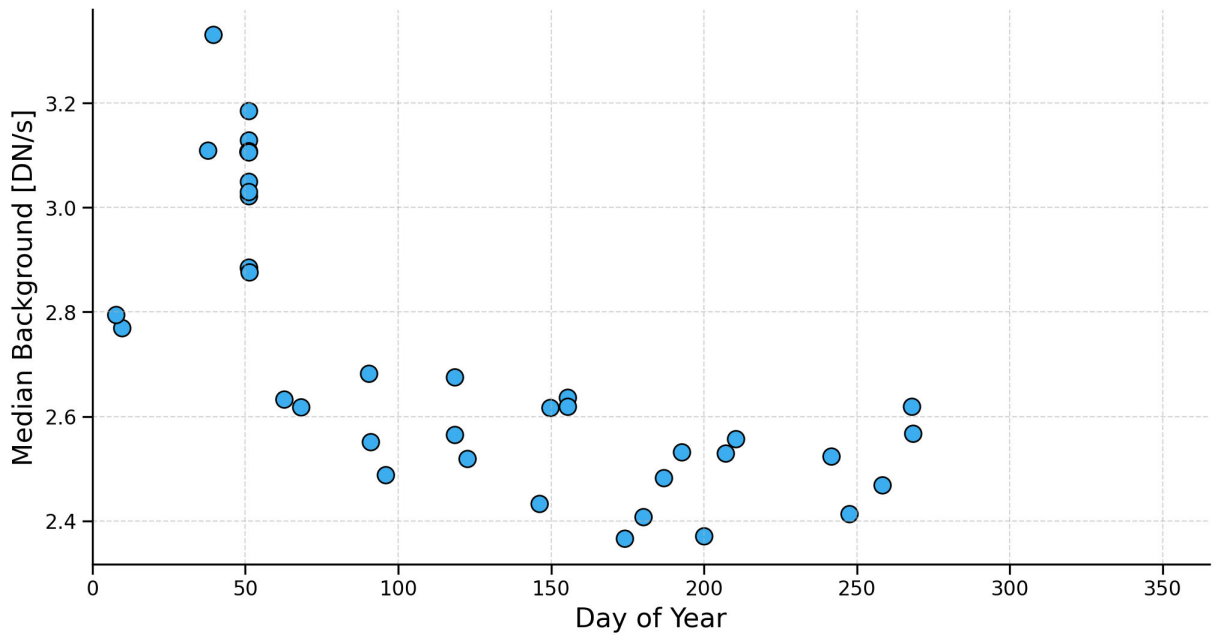

*Figure 8 Median Background levels measured from repeated observations of BD+60 1753 as a function of day of year demonstrating the seasonal effect. Despite the high ecliptic latitude of this target, a clear seasonal trend is observed, with background levels varying by as much as ~43% (~1 DN/s) over the course of the year. The cluster of elevated points around day 50 corresponds to observations from Program 1512, which sampled the pupil wheel to characterize the wavelength solution. These observations exhibit increased variation, likely due to a field star's order 1 trace overlapping with the science target.*

## 7 Quantitative Assessment of Background Subtraction Accuracy

Building on the demonstrated variability in the background levels across different sky positions, we now evaluate the effectiveness of background subtraction using our set of empirically derived templates. For this analysis, we focus on data from science program 2113 (PI: Néstor Espinoza), which includes four, contemporaneous background exposures taken immediately adjacent to the science observations. These background frames are used to reconstruct the background signal to serve as our reference image, providing an optimal baseline for evaluating subtraction accuracy.

Figure 9 illustrates the background subtraction process for one such exposure using the rate data product. The top panel shows the rate image of the original science product, which contains both the dispersed target spectrum and the underlying background. The second panel displays the contemporaneous background reference, followed by the background-subtracted image in the third panel. To better assess the subtraction quality in the regions outside the spectral trace, we apply a source mask and display the masked residuals in the last panel.

Ideally, background subtraction would yield residuals that are statistically consistent with zero in these masked regions. However, in practice, the extending point spread function (PSF) wings and contributions from the blending of multiple spectral orders may leave behind low-level structure that complicates this expectation. At present, we assume these contributions are minimal and that the residuals in the masked regions should be close to zero—an assumption that will be revisited as ongoing efforts, including aperture correction studies and characterizing the spatial profiles of the spectral traces—will help refine these assumptions in the future analyses. In this example, the residuals appear well-behaved, suggesting effective subtraction as we have eliminated the background structure (visually), and with approximately 17% of the image available for background assessment.

To evaluate background subtraction performance of our empirical backgrounds, we compare its results against the existing commissioning reference template and the contemporaneous background constructed from program 2113. The commissioning background, reflects the standard template used in many pipeline reductions. The contemporaneous background serves as a best-case reference, representing the ideal scenario where a dedicated background is taken alongside the science exposures.

Assessing the effectiveness of the background subtraction, we apply each background model using three scaling strategies: (1) no scaling, (2) a one-parameter (global) scaling, and (3) a two-parameter (split-region) scaling approach adopted from the ExoTDRF (formerly supreme-spoon, Radica et al. 2024). The two-parameter scaling method independently scales the left-hand and right-hand sides of the detector relative to the discontinuity of the background. To determine the separation point, we apply a gradient-based edge detection algorithm to the background image iteratively along the rows of the image, which identifies the column where the intensity gradient is maximized—corresponding to the location of the discontinuity. The found columns are then used to divide the image into two regions for separate scaling.

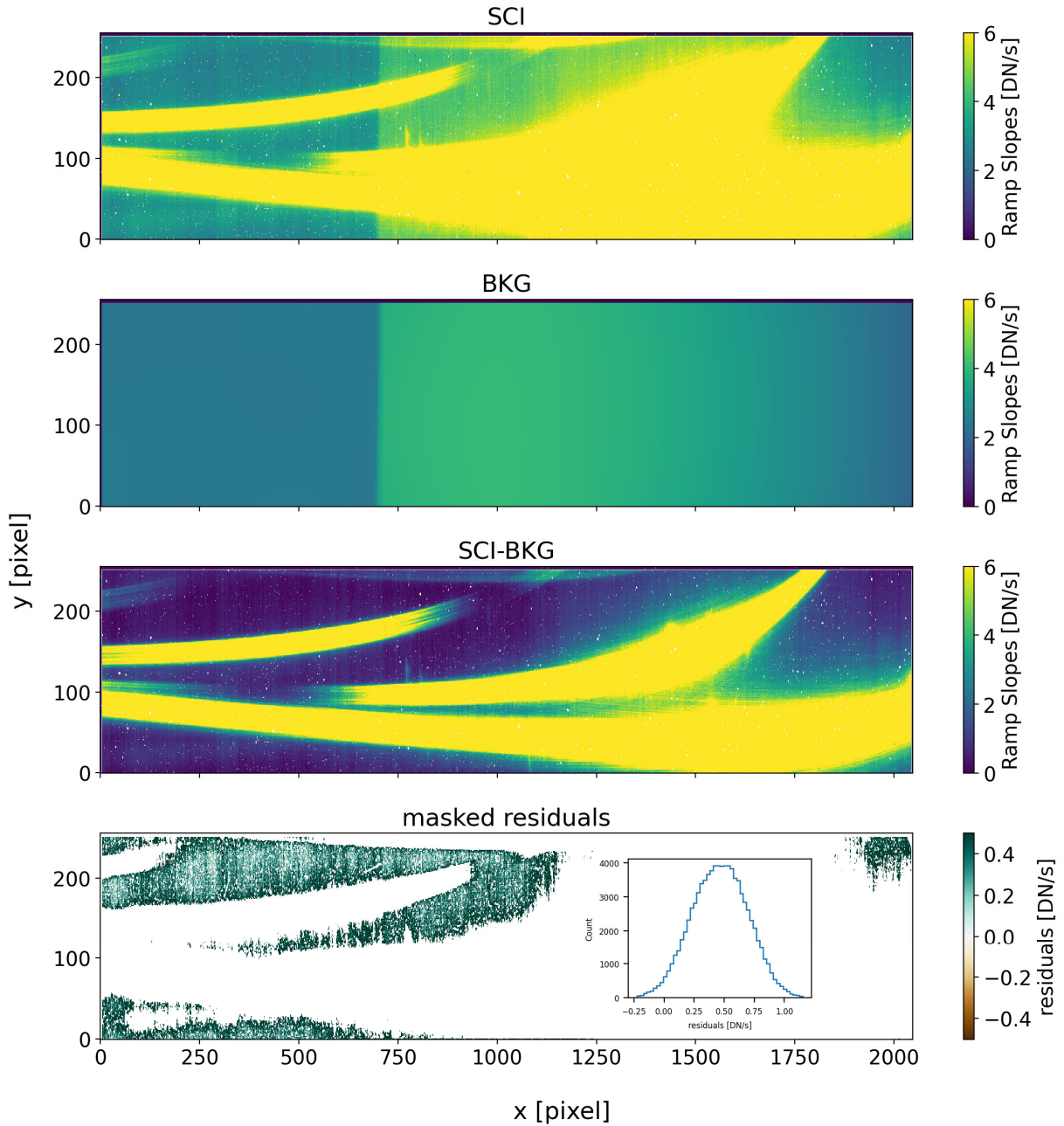

*Figure 9. Background subtraction for a representative exposure from program 2113. Top to third panels: Raw science image (SCI), background template (BKG), and subtraction residuals (SCI–BKG). Bottom: Masked residuals showing background-only pixels (~17%) used to assess subtraction. The histogram shows a near-normal distribution with a mean offset of ~0.46 DN/s, possibly due to residual PSF wings. Nonetheless, the large-scale background is effectively removed.*

Scaling of the background requires measuring the ratio of the flux between one or two distinct regions of the background using either a one-parameter or two-parameter, presumably. For the one-parameter, we chose a region of $x \in [500, 850]$, $y \in [210, 230]$ while for the two-parameter we chose $x \in [500, 680]$, $y \in [205, 230]$ and $x \in [715, 745]$, $y \in [205, 230]$ left and right of the discontinuity, respectively. These methods allow us to quantify how well each background model aligns with the observed data and whether additional adjustments improve subtraction accuracy.

To isolate background-dominated pixels for evaluation, we generate a background mask by applying our source detection algorithm to the original science image, followed by sigma-clipping at $3\sigma$ to exclude extended wings or subtle trace artifacts not captured in the masking step. We then compute two key metrics: (1) The Root Mean Square Error (RMSE) of residuals, which captures the flatness of the background-subtracted image, and (2) the median of the residuals, which measures systematic offsets (i.e., bias) in subtraction—where positive values indicate under-subtraction and negative values indicate over-subtraction.

Figure 10 summarizes the RMSE and median residuals for each background subtraction method across all empirical templates, alongside the contemporaneous and commissioning references, both with and without scaling applied. To support visual inspection, we include residual maps for all three scaling methods in the Appendix see Figures B-1 to B-3, allowing side-by-side comparison of subtraction quality across templates.

Without any scaling, subtraction performance varies widely across empirical templates, with clear signs of both over- and under-subtraction. This spread suggests that while some templates match the spatial structure of the background well, mismatches in intensity introduce significant residual structure. Notably, a subset of templates—such as IDs 4, 8, and 15—perform comparably or better than the contemporaneous benchmark, even without scaling.

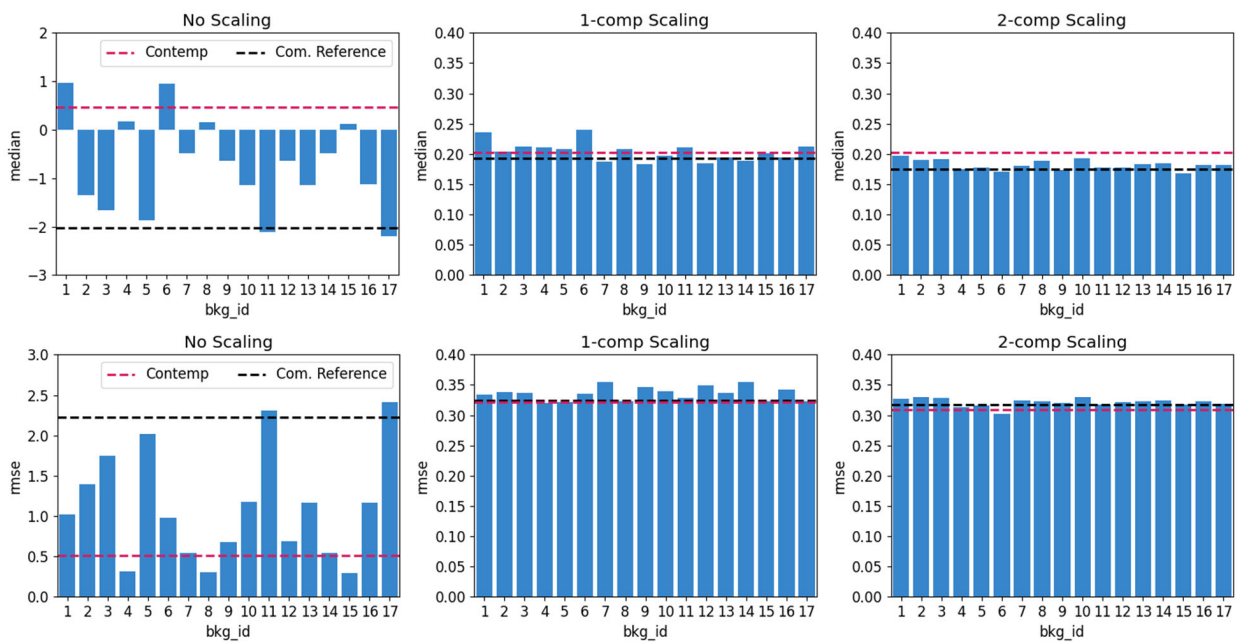

*Figure 10. RMSE and median residuals for 17 empirical backgrounds using three subtraction methods: no scaling (left), single-component (center), and two-component scaling (right). Unscaled subtraction shows high variability and bias, while single-component scaling improves consistency. Two-component scaling performs best, minimizing residuals and correcting intensity mismatches near the spectral discontinuity.*

Introducing single-component scaling significantly improves performance across the board, reducing both variability and overall residuals. RMSE values cluster around those from the scaled contemporaneous and commissioning backgrounds. Importantly, the residual medians become uniformly positive, indicating that the global scaling effectively mitigates over-subtraction seen in the unscaled case.

However, we observe that intensity mismatches remain in localized regions—particularly near the known background discontinuity in the spatial direction—where a single global scaler proves insufficient. The two-component scaling method addresses this by separately balancing the background on either side of the discontinuity. This approach yields the most uniform results, with both RMSEs and medians closely matching the scaled contemporaneous background, demonstrating its effectiveness in modeling spatial intensity breaks.

While global metrics (RMSE, median) show overall improvement with scaling, our inspection of residual maps reveals that some background templates consistently outperform others near the discontinuity, where sharp transitions are most difficult to model. This qualitative difference is not fully captured by global metrics alone. To address this, we propose introducing a localized MSE metric computed specifically near the discontinuity region in future work. This would quantify performance in the most challenging areas and better differentiate between templates.

## 8 Discussion

Qualitatively, the empirical backgrounds show RMSE value approximately 4% higher than the contemporaneous case under the two-parameter scaling method, indicating a moderate increase in pixel-level residual scatter. However, the median residuals for the empirical templates are about 10% lower than those of the contemporaneous background. While this might superficially suggest better subtraction performance, we interpret this result as a potential signature of slight over-subtraction, Specifically, the scaling algorithm may overcompensate in regions near to the spectral trace—particularly in the wings—where faint PSF signal remains after masking. This residual flux can bias the subtraction scaling, especially when using a large number of pixels to estimate the scale factors. The result is a subtraction that removes slightly too much background flux, leading to an inverse dilution effect, where the remaining target flux is underestimated. This can artificially deepen transit depths, especially in cases where the background is comparable to or exceeds the source signal (e.g., during ingress/egress or for fainter targets).

Even in the absence of a strong wavelength-dependent residual pattern, such dilution may still manifest as a wavelength-dependent bias due to column-to-column variations in background structure—particularly near the discontinuity or in lower-signal regions. While precise quantification of this effect would require additional modeling, such as pixel-level trace-to-background flux ratios or forward-modeling of synthetic transits, it remains an important caveat for high-precision spectrophotometry. Investigating this further with faint targets, or with spatially restricted scaling regions that exclude trace-adjacent pixels, may help determine whether a small, constant offset component is systematically present in some templates.

## 9  Conclusions

This work presents a detailed analysis and data-driven framework for characterizing and subtracting the zodiacal light-dominated sky background in NIRISS/SOSS observations. Through simulation and empirical data, we investigate the origin, spatial and seasonal background level variations, and optimal background subtraction to find the following conclusions:

- The GR700XD background arises from the combination of the dispersed spectral orders of the spatially uniform sky background, with a clear discontinuity around pixel column 700 due to the allowable locations of the projected $0^{th}$ order. The overall shape of background remains largely fixed across observations, while absolute intensity varies depending on sky position.

- Empirical backgrounds derived from calibration programs (PIDs 4479 and 6658) capture variability across 17 sky pointings and more will be observed in cycle 4.

- Seasonal monitoring of BD+60 1753 reveals background level fluctuations of up to ~42% for a high ecliptic latitude, driven by seasonal variations and possible bright field star contaminants.

- Background subtraction analysis using our empirical backgrounds and current background template with a test science exposure and contemporaneous background from program 2113 show:

    o Unscaled templates show wide variability in performance, with some significantly over- or under-subtracting the background. While a few cases match or slightly outperform the contemporaneous background in terms of residuals, this may be influenced by systematic over-subtraction effects, as discussed above, and should be interpreted with caution.

    o One- and two-component scaling significantly improve consistency and accuracy.

    o Two-component scaled background yield results within ~4% RMSE and 10% lower median residuals compared to the scaled contemporaneous reference. Users should consider these differences when planning observations to ensure they meet their desired signal-to-noise.

- Residuals near the spectral wings may introduce biases in transit depth measurements if background levels are mis-estimated.

- The 17 empirical background templates provide flexible and viable alternatives to the current default background template. While the contemporaneous background is generally assumed to offer the best match, some empirical templates perform comparably—or better—in localized regions. Our results suggest that even when a contemporaneous background is available, users may benefit from testing empirical templates and applying scaling to optimize subtraction performance.

- The JWST Background Tool suggests that the difference in variability between the left and right-hand side background (where order 0 is important) comes most from the >3μm part of the zodiacal light spectrum. Investigating this model/observation correlation is a promising area of investigation in future studies.

In conclusion, this work represents an important step toward data-driven background modeling for NIRISS/SOSS. Our empirical background library, paired with a flexible scaling framework, provides a practical solution for improving calibration—particularly in programs lacking contemporaneous backgrounds. The full set of background templates, will be made available to the community, as a CRDS SOSS Background reference file, to help users identify the optimal background subtraction strategy for their observations. Efforts are currently underway to integrate a background subtraction step into the official JWST calibration pipeline.

## 11  Appendix

This appendix includes Table A-1 and Figures B-1 through B-3, which offer additional supporting material. These visuals provide additional information and further context for the calibration and background analysis discussed in the main text. The table and figures includes a brief description of relevance.

*Table A-1 Table of the Background Datasets and the corresponding files that went into creating/extracting a background for a given sky pointing.*

| Background ID | FILESET |
|---|---|
| 1 | jw06658007002_02105_00001_nis_rate.fits |
|   | jw06658007002_02101_00001_nis_rate.fits |
|   | jw06658007002_02103_00001_nis_rate.fits |
|   | jw06658007001_04107_00001_nis_rate.fits |
|   | jw06658007001_04101_00001_nis_rate.fits |
|   | jw06658007001_04103_00001_nis_rate.fits |
|   | jw06658007001_04105_00001_nis_rate.fits |
|   | jw06658007003_03101_00001_nis_rate.fits |
|   | jw06658007003_03103_00001_nis_rate.fits |
| 2 | jw06658004003_02103_00001_nis_rate.fits |
|   | jw06658004003_02101_00001_nis_rate.fits |
|   | jw06658004005_02103_00001_nis_rate.fits |
|   | jw06658004005_02101_00001_nis_rate.fits |
|   | jw06658004001_02101_00001_nis_rate.fits |
|   | jw06658004001_02103_00001_nis_rate.fits |
|   | jw06658004002_02101_00001_nis_rate.fits |
|   | jw06658004002_02103_00001_nis_rate.fits |
| 3 | jw04479007002_02101_00001_nis_rate.fits |
|   | jw04479007002_02103_00001_nis_rate.fits |
|   | jw04479007001_02103_00001_nis_rate.fits |
|   | jw04479007001_02101_00001_nis_rate.fits |
|   | jw04479007003_02101_00001_nis_rate.fits |
|   | jw04479007003_02103_00001_nis_rate.fits |
| 4 | jw06658005003_02101_00001_nis_rate.fits |
|   | jw06658005003_02103_00001_nis_rate.fits |
|   | jw06658005002_02101_00001_nis_rate.fits |
|   | jw06658005002_02107_00001_nis_rate.fits |
|   | jw06658005002_02103_00001_nis_rate.fits |
|   | jw06658005002_02105_00001_nis_rate.fits |
| 5 | jw04479008004_02101_00001_nis_rate.fits |
|   | jw04479008002_02101_00001_nis_rate.fits |
|   | jw04479008003_02101_00001_nis_rate.fits |
|   | jw04479008001_02101_00001_nis_rate.fits |
|   | jw04479008010_02101_00001_nis_rate.fits |
|   | jw04479008006_02101_00001_nis_rate.fits |
|   | jw04479008005_02101_00001_nis_rate.fits |
|   | jw04479008007_02101_00001_nis_rate.fits |
| | jw04479008009_02101_00001_nis_rate.fits |
| | jw04479008008_02101_00001_nis_rate.fits |
| 6 | jw06658008002_02103_00001_nis_rate.fits |
| | jw06658008002_02101_00001_nis_rate.fits |
| | jw06658008002_02107_00001_nis_rate.fits |
| | jw06658008002_02105_00001_nis_rate.fits |
| | jw06658008001_03105_00001_nis_rate.fits |
| | jw06658008001_03107_00001_nis_rate.fits |
| | jw06658008001_03103_00001_nis_rate.fits |
| | jw06658008001_03101_00001_nis_rate.fits |
| | jw06658008003_02101_00001_nis_rate.fits |
| 7 | jw06658001001_02101_00001_nis_rate.fits |
| | jw06658001009_02101_00001_nis_rate.fits |
| | jw06658001003_02101_00001_nis_rate.fits |
| | jw06658001006_02101_00001_nis_rate.fits |
| | jw06658001004_02101_00001_nis_rate.fits |
| | jw06658001005_02101_00001_nis_rate.fits |
| | jw06658001002_02101_00001_nis_rate.fits |
| | jw06658001008_02101_00001_nis_rate.fits |
| | jw06658001007_02101_00001_nis_rate.fits |
| | jw06658001010_02101_00001_nis_rate.fits |
| 8 | jw06658009005_02101_00001_nis_rate.fits |
| | jw06658009005_02103_00001_nis_rate.fits |
| | jw06658009002_02103_00001_nis_rate.fits |
| | jw06658009002_02101_00001_nis_rate.fits |
| | jw06658009003_02101_00001_nis_rate.fits |
| | jw06658009001_02103_00001_nis_rate.fits |
| | jw06658009001_02101_00001_nis_rate.fits |
| | jw06658009004_02101_00001_nis_rate.fits |
| | jw06658009004_02103_00001_nis_rate.fits |
| 9 | jw04479001009_03101_00001_nis_rate.fits |
| | jw04479001002_03101_00001_nis_rate.fits |
| | jw04479001004_03101_00001_nis_rate.fits |
| | jw04479001007_03101_00001_nis_rate.fits |
| | jw04479001003_03101_00001_nis_rate.fits |
| | jw04479001005_03101_00001_nis_rate.fits |
| | jw04479001006_03101_00001_nis_rate.fits |
| | jw04479001008_03101_00001_nis_rate.fits |
| | jw04479001001_03101_00001_nis_rate.fits |
| | jw04479001010_03101_00001_nis_rate.fits |
| 10 | jw04479005004_02101_00001_nis_rate.fits |
| | jw04479005004_02103_00001_nis_rate.fits |
| | jw04479005003_02103_00001_nis_rate.fits |
| | jw04479005003_02101_00001_nis_rate.fits |
| | jw04479005001_02101_00001_nis_rate.fits |
| | jw04479005001_02103_00001_nis_rate.fits |
| | jw04479005005_02103_00001_nis_rate.fits |
| | jw04479005005_02101_00001_nis_rate.fits |
| | jw04479005002_02103_00001_nis_rate.fits |
| | jw04479005002_02101_00001_nis_rate.fits |
| 11 | jw04479007005_02101_00001_nis_rate.fits |
| | jw04479007005_02103_00001_nis_rate.fits |
| | jw04479007004_02101_00001_nis_rate.fits |
| | jw04479007004_02103_00001_nis_rate.fits |
| 12 | jw04479002001_03101_00001_nis_rate.fits |
| | jw04479002007_03101_00001_nis_rate.fits |
| | jw04479002006_03101_00001_nis_rate.fits |
| | jw04479002008_03101_00001_nis_rate.fits |
| | jw04479002004_03101_00001_nis_rate.fits |
| | jw04479002005_03101_00001_nis_rate.fits |
| | jw04479002003_03101_00001_nis_rate.fits |
| | jw04479002009_03101_00001_nis_rate.fits |
| | jw04479002002_03101_00001_nis_rate.fits |
| | jw04479002010_03101_00001_nis_rate.fits |
| 13 | jw04479003001_02103_00001_nis_rate.fits |
| | jw04479003001_02101_00001_nis_rate.fits |
| | jw04479003002_02101_00001_nis_rate.fits |
| | jw04479003002_02103_00001_nis_rate.fits |
| | jw04479003003_02101_00001_nis_rate.fits |
| | jw04479003003_02103_00001_nis_rate.fits |
| | jw04479003005_02103_00001_nis_rate.fits |
| | jw04479003005_02101_00001_nis_rate.fits |
| | jw04479003004_02101_00001_nis_rate.fits |
| | jw04479003004_02103_00001_nis_rate.fits |
| 14 | jw06658003010_02101_00001_nis_rate.fits |
| | jw06658003008_02101_00001_nis_rate.fits |
| | jw06658003006_02101_00001_nis_rate.fits |
| | jw06658003001_02101_00001_nis_rate.fits |
| | jw06658003003_02101_00001_nis_rate.fits |
| | jw06658003009_02101_00001_nis_rate.fits |
| | jw06658003004_02101_00001_nis_rate.fits |
| | jw06658003005_02101_00001_nis_rate.fits |
| | jw06658003002_02101_00001_nis_rate.fits |
| | jw06658003007_02101_00001_nis_rate.fits |
| 15 | jw06658005001_02105_00001_nis_rate.fits |
| | jw06658005001_02103_00001_nis_rate.fits |
| | jw06658005001_02101_00001_nis_rate.fits |
| | jw06658005001_02107_00001_nis_rate.fits |
| 16 | jw04479006001_02107_00001_nis_rate.fits |
| | jw04479006001_02101_00001_nis_rate.fits |
| | jw04479006001_02103_00001_nis_rate.fits |
| | jw04479006001_02105_00001_nis_rate.fits |
| | jw04479006003_02103_00001_nis_rate.fits |
| | jw04479006003_02101_00001_nis_rate.fits |
| | jw04479006002_02101_00001_nis_rate.fits |
| | jw04479006002_02103_00001_nis_rate.fits |
| | jw04479006002_02105_00001_nis_rate.fits |
| | jw04479006002_02107_00001_nis_rate.fits |
| **17** | jw01541005001_0210p_00001_nis_rate.fits |
| | jw01541005001_0210j_00001_nis_rate.fits |
| | jw01541005001_0210l_00001_nis_rate.fits |
| | jw01541005001_02107_00001_nis_rate.fits |
| | jw01541005001_0210t_00001_nis_rate.fits |
| | jw01541005001_02103_00001_nis_rate.fits |
| | jw01541005001_02109_00001_nis_rate.fits |
| | jw01541005001_0210x_00001_nis_rate.fits |
| | jw01541005001_0210z_00001_nis_rate.fits |
| | jw01541005001_0210h_00001_nis_rate.fits |
| | jw01541005001_02101_00001_nis_rate.fits |
| | jw01541005001_0210n_00001_nis_rate.fits |
| | jw01541005001_02105_00001_nis_rate.fits |
| | jw01541005001_0210r_00001_nis_rate.fits |
| | jw01541005001_0210v_00001_nis_rate.fits |
| | jw01541005001_0210b_00001_nis_rate.fits |
| | jw01541005001_0210d_00001_nis_rate.fits |
| | jw01541005001_0210f_00001_nis_rate.fits |

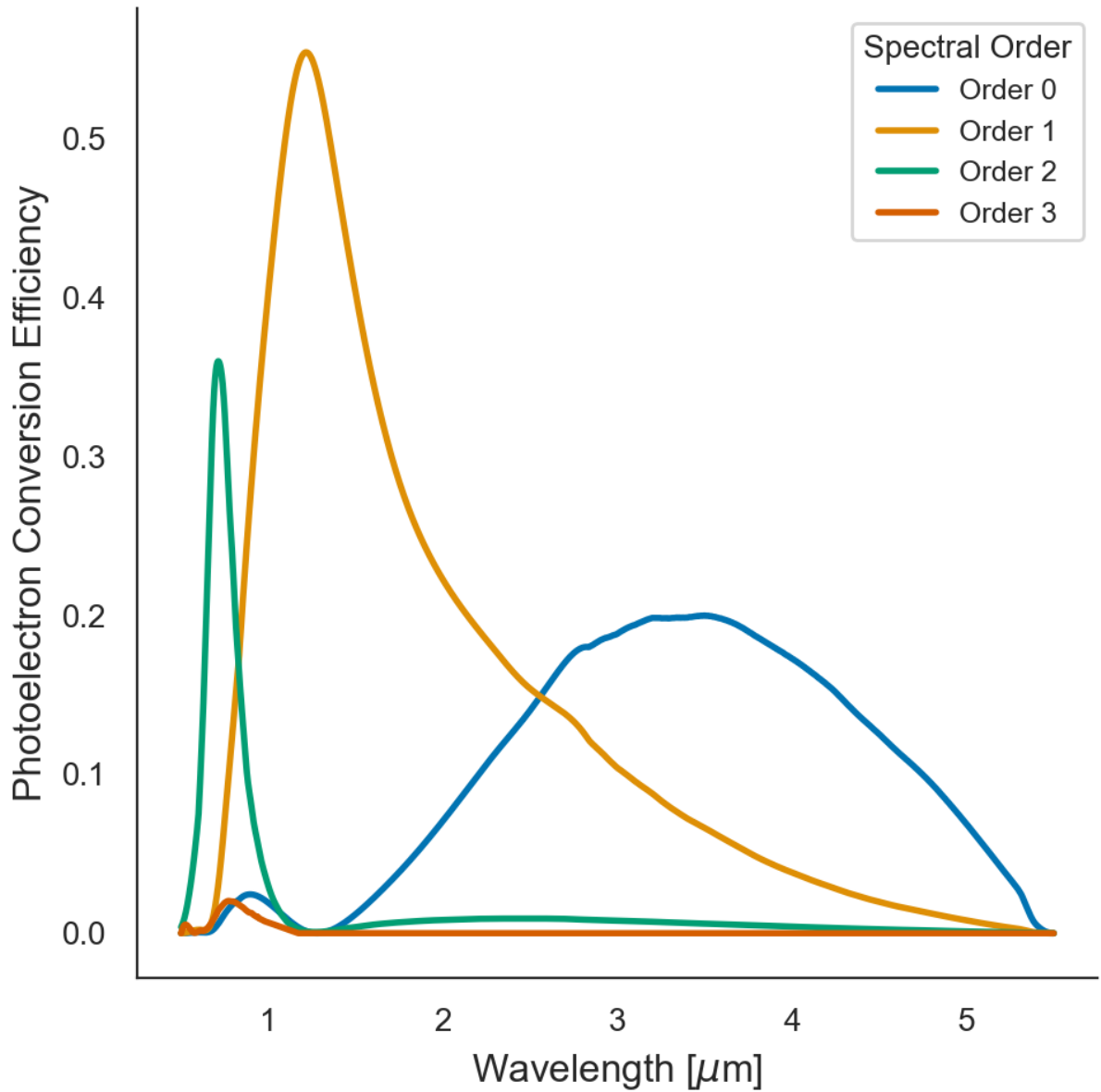

*Figure A-1: Photoelectron Conversion efficiency (PCE) curves for spectral order 0, 1, 2, 3. The zeroth-order PCE was derived during ground testing while order 1 to 3 are the most recent profiles. It's clear that the order 0 throughput is quite broad peaking at longer wavelengths relative the other spectral order which are narrower and peak at shorter wavelength (below 2 µm).*

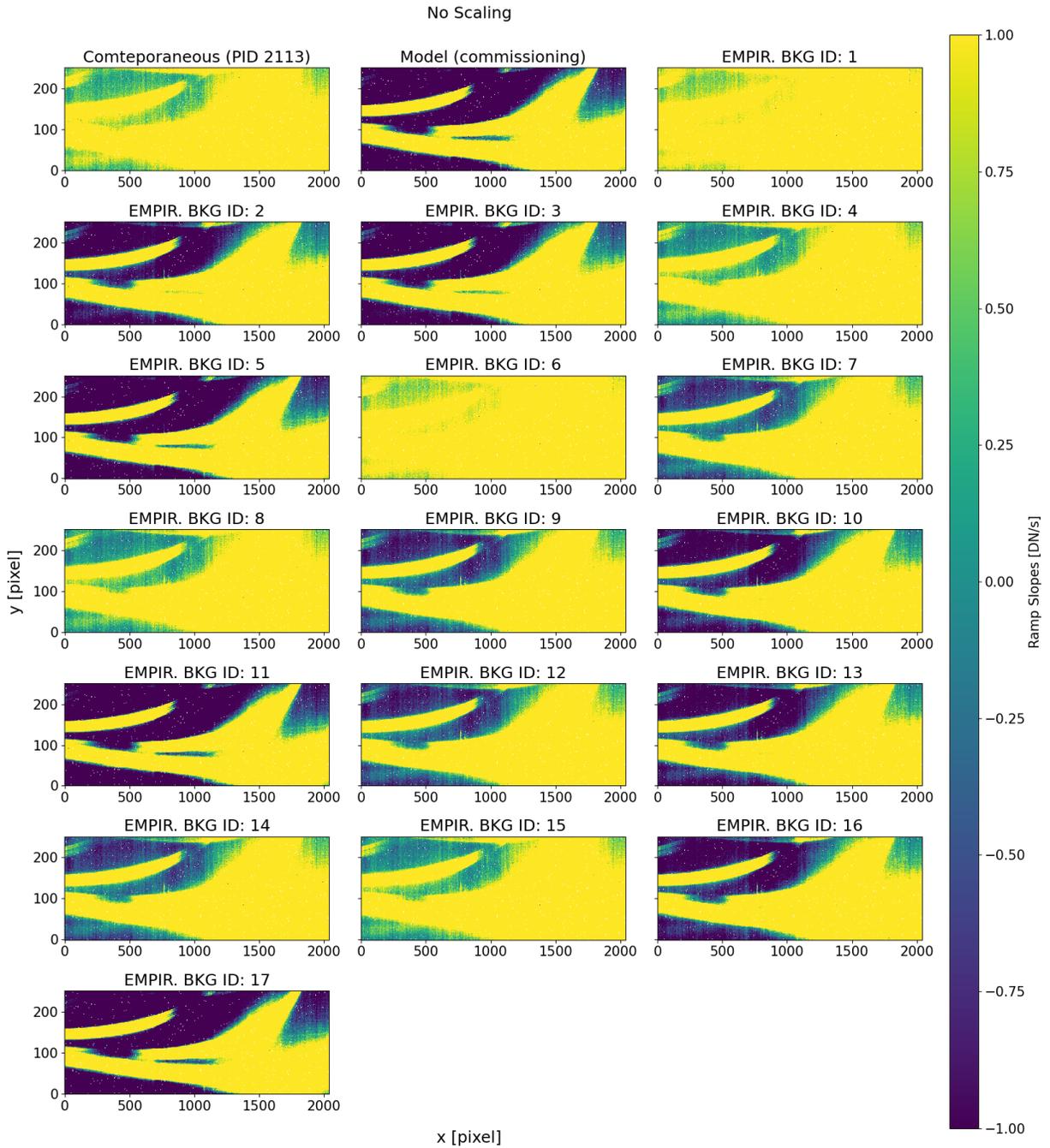

*Figure B-1 Background-subtracted images using the unscaled backgrounds. A set of the background are over-subtracted and is noticeable around the discontinuity (around pixel column x=700). In under-subtracted cases most of the background structure is removed, however, a residual offset persists.*

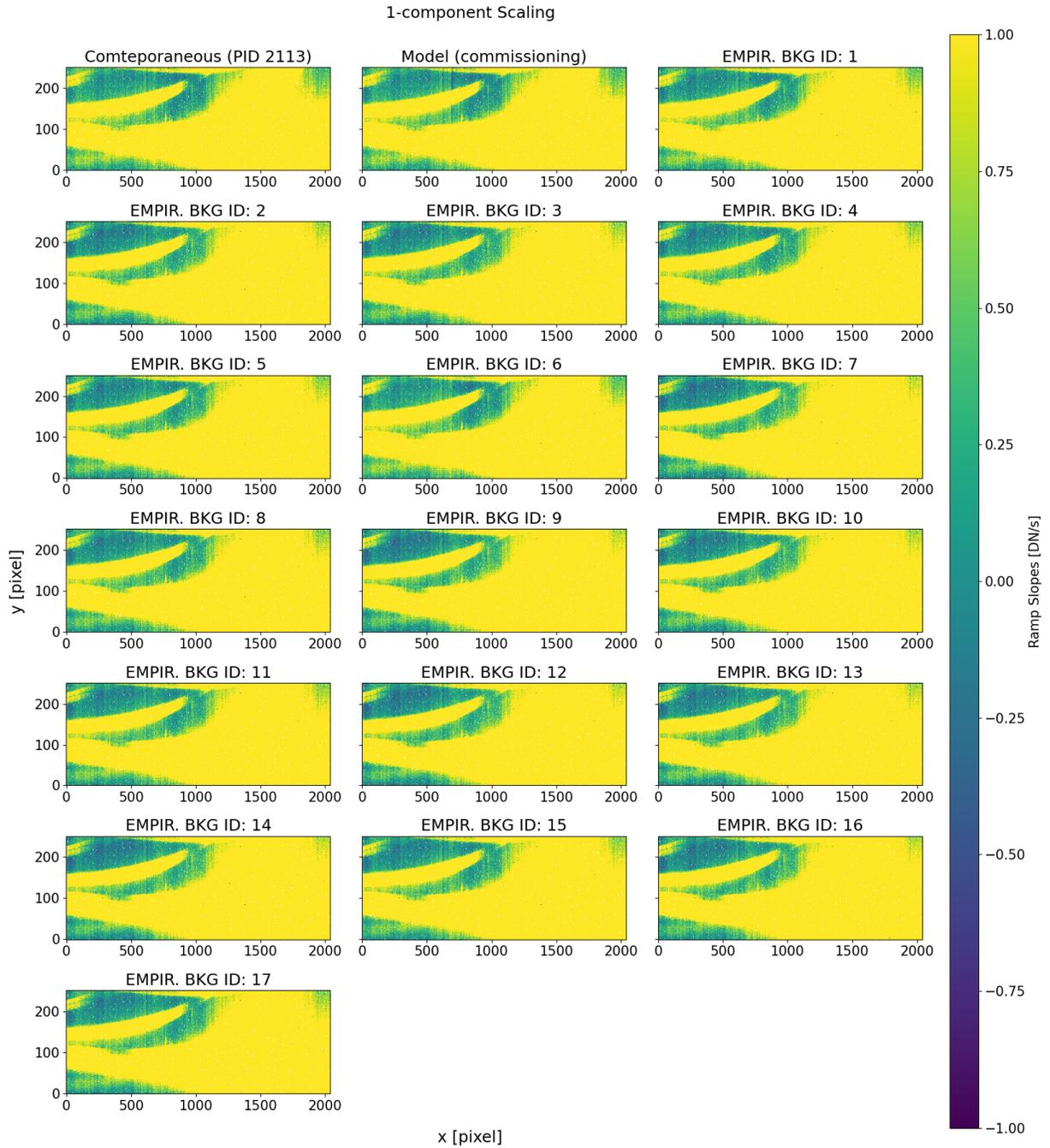

*Figure B-2 Background-subtracted images using the one-parameter "global" scaled backgrounds. Residual structure remains around the discontinuity (pixel column x=700) across many of the scaled background including the contemporaneous template.*

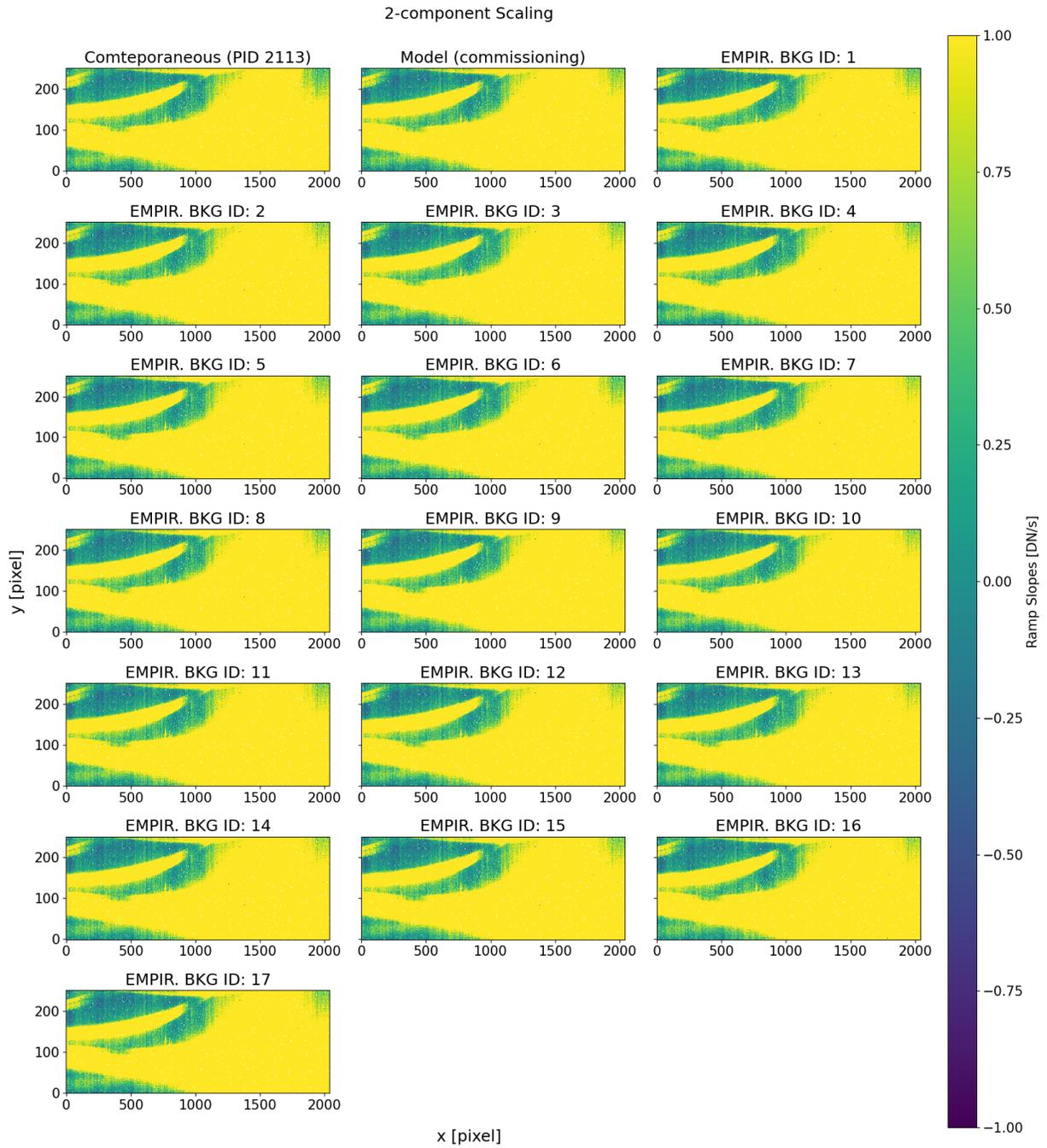

*Figure B-3 Background-subtracted images using the two-parameter "split-region" scaled backgrounds. Visual inspection suggests improved residual structure compared to one-parameter scaling.*